\documentclass[aps,prd,preprint,floats,showpacs,superscriptaddress]{revtex4}
\usepackage{graphicx}
\begin{document}

\setcounter{page}{0}
\thispagestyle{empty}

\preprint{$
\begin{array}{l}
\mbox{ANL-HEP-PR-04-29}\\
\mbox{NSF-KITP-04-39}\\
\mbox{hep-ph/0404158}\\[5mm]
\end{array}
$}
\vspace*{2cm}

\title{\bf Transverse momentum distribution of $\Upsilon$ 
  production \\ in hadronic collisions}

\author{Edmond~L.~Berger}
\email[e-mail: ]{berger@anl.gov}
\affiliation{High Energy Physics Division, 
Argonne National Laboratory, Argonne, IL 60439}
\author{Jianwei~Qiu}
\email[e-mail: ]{jwq@iastate.edu}
\affiliation{Department of Physics and Astronomy, 
Iowa State University, Ames, IA 50011}
\author{Yili~Wang}
\email[e-mail: ]{yiliwa@iastate.edu}
\affiliation{Department of Physics and Astronomy, 
Iowa State University, Ames, IA 50011}

\date{April 19, 2004}

\begin{abstract}
We calculate the transverse momentum $p_T$ distribution for production 
of the $\Upsilon$ states in hadronic reactions.  For small 
$p_T (\leq M_\Upsilon)$, we resum to all orders in the strong coupling 
$\alpha_s$ the process-independent large logarithmic contributions that 
arise from initial-state gluon showers.  We demonstrate that the  
$p_T$ distribution at low $p_T$ is dominated by the region of small 
impact parameter $b$ and that it may be computed reliably in perturbation 
theory.  We express the cross section at large $p_T$ by the 
${\cal O} (\alpha_s^3)$ lowest-order non-vanishing perturbative contribution. 
Our results are consistent with data from the Fermilab Tevatron collider. 

\end{abstract}


\maketitle

\section{Introduction}
\label{sec:intro}

The theoretical description of the transverse momentum $p_T$
distribution of heavy quarkonium 
production in hadron
collisions raises interesting challenges.  Most calculations within 
the framework of perturbative quantum chromodynamics (QCD) consider 
the distribution at large $p_T$ at collider energies and tend not to 
address the region of low $p_T$ where the cross section is greatest and 
the bulk of the data lie~\cite{Cho:1995vh,Amundson:1996qr,reviews}.  A purely 
phenomenological fit to the low $p_T$ data on $\Upsilon$ 
production~\cite{cdf2} appears to require sizable non-perturbative 
parton-$k_T$ smearing~\cite{Schuler:1996ku}.  From a theoretical point 
of view, the region of low $p_T$ is expected to be influenced strongly by 
initial-state gluon showering.  
A fixed-order perturbative treatment in QCD leads to singular terms 
in the region of small $p_T$ of the type $1/p_T^2$, enhanced by large 
higher-order logarithmic contributions caused by initial-state gluon 
radiation.  These contributions have the form 
$\alpha_s\log^2(M_{\Upsilon}^2/p_T^2)$ for every power of the strong 
coupling $\alpha_s$, and reliable predictions, especially in the
regions of small and moderate $p_T$, require that the logarithmic
contributions be summed to all orders in $\alpha_s$.

The Collins-Soper-Sterman (CSS) impact parameter $b$-space resummation 
formalism~\cite{Collins:1984kg} has been used successfully for the
all-orders resummation of large initial-state logarithmic terms in
several cases of physical  
interest~\cite{Davies:1984sp,Qiu:2000hf,Balazs:1997hv,Berger:yp,Catani:vd,Catani:2000vq,Berger:2002ut,Kulesza:2003wi}.
In this paper, we argue and demonstrate that the resummation formalism
should apply at the scale of the $\Upsilon$ mass in hadronic collisions
at collider energies.  We extend the formalism and use it to compute
the $p_T$ distribution of the $\Upsilon$ states~\cite{fixed}.
We obtain good agreement with the  
data~\cite{cdf2,Acosta:2001gv} on the $p_T$ distribution of $\Upsilon$
production at Tevatron energies for all $p_T$.  

Different from the production of the $W$, $Z$, and Higgs bosons, or of a 
virtual photon in the Drell-Yan process, the $\Upsilon$ is unlikely 
to be produced in pointlike fashion in a short-distance hard collision. 
Instead, a bottom quark $b\bar{b}$ pair is produced in the hard collision 
and then transmutes into a colorless $\Upsilon$ meson through soft radiation 
and coherent self-interaction.  Therefore, there are issues to address before 
the CSS formalism can be applied to $\Upsilon$ production.  These include 
the color structure of the lowest order production processes: 
$q\bar{q}\rightarrow b\bar{b}(Q)$ and $gg\rightarrow  b\bar{b}(Q)$, and 
the relatively small value of the $b \bar{b}$ pair mass $Q$.
   
Most applications of the resummation formalism are to the production
of systems that are color singlets whereas the $b \bar{b}$ system
produced in $q\bar{q}\rightarrow b\bar{b}$ 
and $gg\rightarrow  b\bar{b}$ need not be color neutral.  Nevertheless, 
because the $b$ quark mass is large, gluon radiation is suppressed from the 
final-state heavy quark lines and from virtual exchange 
lines that lead to the production of heavy quark pairs~\cite{Berger:yp}.  
Correspondingly, the important logarithmic terms are associated with  
gluon radiation from the active initial-state partons, the same as those 
in massive lepton-pair (Drell-Yan) and Higgs boson production.  The 
process-independent leading logarithmic terms do  
not depend on the color of the heavy quark pair.  Color dependence becomes 
relevant for the higher order terms, as explained in Sec.~\ref{sec:prodsmall}. 

The overall center-of-mass energy $\sqrt{S}$ dependence of the CSS $b$-space 
distribution function is examined by Qiu and Zhang in Ref.~\cite{Qiu:2000hf}.  
They show that the location of the saddle point of this distribution 
can be well within the perturbative region of small $b$ for $Q$ as small as
6~GeV at the Tevatron collider energy.  The resummed $b$-space distribution is 
peaked strongly in the perturbative region of small $b$, as we show in 
Sec.~\ref{sec:prodsmall}, and the $p_T$ distribution of $\Upsilon$ production 
should be amenable to a resummation treatment.  Despite the fact that the 
logarithmic term $\ln Q$ is not large, the large value of $\sqrt{S}$ opens 
a large region of phase space for gluon emission.  Correspondingly, as is 
demonstrated in this paper, the shape of the $p_T$ distribution for 
$\Upsilon$ production is determined by the resummable part of the gluon shower 
and is predictable quantitatively at low $p_T$.  

We begin in Sec.~\ref{sec:prod} with the basic assumption that the $p_T$  
distribution of $\Upsilon$ production is derived from the $p_T$ distribution 
for the production of a pair $b \bar{b}$ of bottom quarks.  We express the 
differential cross section in terms of a two-step factorization procedure.  
We present our fixed-order perturbative calculation applicable at large 
transverse momentum in Sec.~\ref{sec:fixed} where we also describe models 
that specify the manner in which the $b \bar{b}$ pair transforms into the 
$\Upsilon$.  
In Sec.~\ref{sec:prodsmall}, we specialize to the situation at small $p_T$  
and summarize the required parts of the all-orders resummation formalism.  
Section~\ref{sec:results} is devoted to our numerical results 
and comparison with data.  We provide of our conclusions and discuss 
potential improvements of our calculation in Sec.~\ref{sec:conclusions}.

\section{Production Dynamics}
\label{sec:prod}

We use a two-step factorization procedure to 
represent production of the $\Upsilon$ states, with particular attention 
to the prediction of transverse momentum distributions. 
We begin with the assumption that a pair of bottom quarks $b \bar{b}$
is produced  
in a hard-scattering short-distance process:
\begin{equation}
A(p_A)+B(p_B)\rightarrow b\bar{b}(Q)[\rightarrow \Upsilon(p)+\bar{X}]
+ X'\, .
\end{equation}
Because the mass $Q$ of the $b \bar{b}$ pair is large,  
the pair is produced at a distance {scale $\sim 1/(2m_b) \sim 1/45$~fm}.  
This scale is much smaller than the physical size of a $\Upsilon$ meson.  The 
compact $b\bar{b}$ pair may represent the minimal Fock state of the
$\Upsilon$,  
but the overlap of this minimal Fock state with the full wave function of the 
$\Upsilon$ is perhaps small, as is suggested by the inadequacies of the 
color-singlet approach~\cite{Berger:1980ni} in some situations~\cite{reviews}, 
and other components of the wave-function must be considered.  Alternatively, 
one may realize that the compact $b\bar{b}$ system is unlikely to become an 
$\Upsilon$ meson at the production point.  Instead, the pair must
expand, and the  
$b$ and $\bar{b}$ will interact with each other coherently until they
transmute  
into a physical $\Upsilon$ meson.  

Once produced in the hard-scattering, a $b\bar{b}$ pair of invariant
mass $Q > 2M_B$  
is more likely to become a 
pair of $B$ mesons.  Therefore, the virtuality of the intermediate
$b\bar{b}$ pair  
should be limited if an $\Upsilon$ is to result.  This limitation of
the virtuality  
allows us to use perturbative factorization and to write the
differential cross  
section in the usual 
way~\cite{Collins:gx,QiuSterman:QQfac}.  For $p_T\gg 2(M_B- m_b)$, we write  
\begin{equation}
\frac{d\sigma_{AB\rightarrow\Upsilon X}}{dp_T^2 dy}
= \sum_{a,b} \int dx_a\, \phi_{a/A}(x_a)\, dx_b\, \phi_{b/B}(x_b)\, 
\frac{d\hat{\sigma}_{ab\rightarrow\Upsilon X}}{dp_T^2 dy}\, .
\label{hadron:xsec}
\end{equation}
In Eq.~(\ref{hadron:xsec}), $p_T$ and $y$ are the transverse momentum
and rapidity  
of the final 
$\Upsilon$.  The functions $\phi_i(x)$ are parton distribution
functions; $x_a$ and 
$x_b$ are fractional light-cone momenta carried by the incident partons; and 
Eq.~(\ref{hadron:xsec}) expresses initial-state collinear factorization.  
The spectator interactions between the beam remnants and the formation of the 
$\Upsilon$ meson are suppressed by one or more powers of $1/p_T^2$.

\begin{figure}[h]
\centerline{\includegraphics[width=10cm]{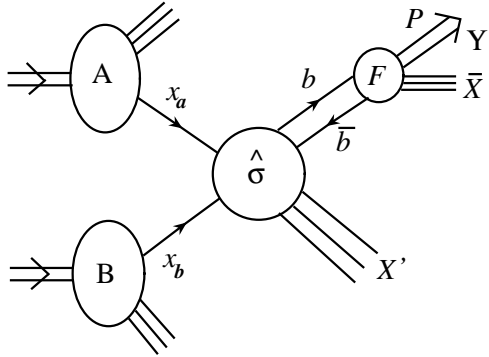}}
\caption{Hadronic production of an $\Upsilon$ via an intermediate 
heavy quark pair $b$ and $\bar{b}$.}
\label{fig1}
\end{figure}

Since the momentum of heavy quark $b$ ($\bar{b}$) in the pair's rest frame is 
much less than the mass of the pair, $Q - 2m_b < 2M_B - 2m_b \ll 2m_b$, the 
parton-level production cross section 
$d\hat{\sigma}_{ab\rightarrow\Upsilon X}/dp_T^2 dy$ in Eq.~(\ref{hadron:xsec}) 
might be factored further~\cite{QiuSterman:QQfac}, as is 
sketched in Fig.~\ref{fig1}.  The incident partons labeled $x_a$ and
$x_b$ interact  
inclusively to produce an off-shell $b \bar{b}$ system plus state
$X'$.  In turn,  
the $b \bar{b}$ system evolves into the $\Upsilon$ plus a system
labeled $\bar{X}$;  
$X = X' + \bar{X}$
This second factored expression is 
\begin{equation}
\frac{d\hat{\sigma}_{ab\rightarrow\Upsilon X}}{dp_T^2 dy}
\approx \sum_{[b\bar{b}]} 
\int dQ^2 
\left[\frac{d\hat{\sigma}_{ab\rightarrow[b\bar{b}](Q) X'}}
           {dQ^2dp_T^2 dy}
\right]
{\cal F}_{[b\bar{b}]\rightarrow\Upsilon \bar{X}}(Q^2)\, .
\label{parton:xsec}
\end{equation}
In writing Eq.~(\ref{parton:xsec}), we approximate the transverse momentum 
and rapidity of the $b\bar{b}$ pair by $p_T$ and $y$, 
respectively, because $Q^2-4m_b^2\ll p_T^2$.

The function $d\hat{\sigma}_{ab\rightarrow[b\bar{b}](Q) X'}/dQ^2dp_T^2dy$
represents a partonic short-distance hard-part for inclusive production of a 
$b\bar{b}$ pair of invariant mass $Q$ and quantum numbers $[b\bar{b}]$. 
This short-distance hard-part is calculable in perturbation theory with the 
parton momenta of all light partons off-mass-shell by at least 
${\rm{min}}(4m_b^2, p_T^2)$.
The function ${\cal F}_{[b\bar{b}]\rightarrow\Upsilon \bar{X}}(Q^2)$
represents  
a transition probability distribution for a $b\bar{b}$ pair of
invariant mass $Q$ and quantum numbers $[b\bar{b}]$ to transmute into
an $\Upsilon$ meson.  It includes all dynamical $b\bar{b}$ interactions 
of momentum scale less than $Q^2-4m_b^2$.  Different assumptions and choices 
for the transition probability distribution ${\cal F}(Q^2)$ lead to different 
models of quarkonium production.  We return to the topic of these models 
in Sec.~\ref{sec:fixed}.

The basic assumptions of this section imply that the transverse momentum 
distributions of the $\Upsilon$ states at transverse momenta 
$p_T \sim M_{\Upsilon}$ 
will reflect the shape of the transverse momentum distribution for production 
of a $b \bar{b}$ pair whose mass $Q \sim M_{\Upsilon}$.  In this paper, we 
focus on the region below $p_T \sim M_{\Upsilon}$.
If $p_T^2\gg Q^2$, all final-state logarithmic terms of the form 
$(\alpha_s \log(p_T^2/Q^2))^N$ can be resummed perturbatively to all orders 
in $\alpha_s$~\cite{Berger:2001wr}.  

\section{FIXED ORDER CALCULATION:\ $p_T\sim M_{\Upsilon}$}
\label{sec:fixed}
 
When the transverse momentum,  
$p_T \sim {\cal O}(M_{\Upsilon})$, 
the collinear factorized expression in 
Eqs.~(\ref{hadron:xsec}) and (\ref{parton:xsec}) remains reliable 
with the partonic short-distance hard parts in 
Eq.~(\ref{parton:xsec}) computed as a power series in $\alpha_s$ in QCD 
perturbation theory.   

The transition probability distribution ${\cal F}(Q^2)$ is introduced 
in Sec.~\ref{sec:prod}.  Different assumptions and 
choices for ${\cal F}(Q^2)$ correspond to different models of quarkonium
production.  In the color evaporation (or color-bleaching) model 
(CEM)~\cite{Amundson:1996qr,CEMearly}, an assumption is made, based 
qualitatively on semi-local duality, that one may safely ignore the 
details of the formation of color-neutral bound states with specific 
quantum numbers $J^{PC}$.  In particular, in the case of states such as 
the $J/\psi$ and $\Upsilon$ that have $J^{PC} = 1^{--}$, 
soft gluon effects are presumed to take care of whatever quantum numbers have 
to be arranged.  Within our framework, this model is effectively represented 
by the statement that 
\begin{equation}
{\cal F}_{[b\bar{b}]\rightarrow\Upsilon}(Q^2)
= \left\{ \begin{array}{lll}
          C_{\Upsilon}  & {\hskip 0.2in} 
                        &  \mbox{if}\quad  
                           4m_b^2 \leq Q^2 \leq 4M_B^2 \\
          0             &    
                        &\mbox{otherwise} \, .
          \end{array}
  \right.
\label{F:CEM}
\end{equation}
The non-perturbative constant $C_{\Upsilon}$ sets the overall 
normalization of the cross section.  Its value cannot be predicted.  It  
changes with the specific state of the 
$\Upsilon$ meson. In the CEM model, the parton-level $\Upsilon$ cross 
section in Eq.~(\ref{parton:xsec}) can be written as
\begin{equation}
\frac{d\hat{\sigma}^{\rm{CEM}}_{ab\rightarrow\Upsilon X}}{dp_T^2 dy}
\approx C_{\Upsilon}
\int_{4m_b^2}^{4M_B^2} dQ^2 
\left[\frac{d\hat{\sigma}_{ab\rightarrow b\bar{b}(Q)}}
           {dQ^2dp_T^2 dy}
\right] \, ,
\label{parton:xsec:CEM}
\end{equation}
where the $b\bar{b}$ final-state includes a sum over all possible quantum
states $[b\bar{b}]$ of the $b\bar{b}$ pair.

In the color singlet model for quarkonium production~\cite{Berger:1980ni}, 
a projection operator is used to place the $b \bar{b}$ system in the 
spin-state of the $\Upsilon$, and explicit gluon radiation guarantees 
charge conjugation (C) and color conservation 
at the level of the hard-scattering amplitude.  The distribution 
${\cal F}_{[b\bar{b}]\rightarrow\Upsilon}(Q^2)$ is proportional to the square 
of the momentum-space wave function of the $\Upsilon$, 
$|\tilde{\Psi}(q)|^2$, with the relative momentum of the $b\bar{b}$
pair $q^2=Q^2-4m_b^2$.  Because the $\Upsilon$ wave function falls steeply, 
one can approximate $Q^2\approx 4m_b^2$ in the $b\bar{b}$ partonic
cross section. 
The integration $\int dQ^2 {\cal
  F}_{[b\bar{b}]\rightarrow\Upsilon}(Q^2)$ 
in Eq.~(\ref{parton:xsec}) leads to  
the square of the $\Upsilon$ wave function at the origin $|\Psi(0)|^2$.

The non-relativistic QCD model (NRQCD)~\cite{Cho:1995vh,Caswell:1985ui} takes 
into consideration that the velocity of the heavy quark $b$ ($\bar{b}$) in the 
rest frame of the $b\bar{b}$ pair is much less than the speed of light.  The 
velocity expansion translates into statements that the distribution 
${\cal F}(Q^2)$ is a steeply falling function of the relative heavy quark 
momentum, $q^2\equiv Q^2-4m_b^2$, and that its moments satisfy the
inequalities   
\begin{equation}
\langle (q^2)^N \rangle
\equiv \int dQ^2\, (q^2)^N\, 
       {\cal F}_{[b\bar{b}]\rightarrow\Upsilon}(Q^2)
\ll (4m_b^2)^N \, ,
\label{q2:moments}
\end{equation}
for moments, $N\ge 1$.  Correspondingly, one can expand the partonic hard 
part in Eq.(\ref{parton:xsec}) at $Q^2=(2m_b)^2$ and obtain 
\begin{equation}
\frac{d\hat{\sigma}^{\rm{NRQCD}}_{ab\rightarrow\Upsilon X}}{dp_T^2 dy}
\approx \sum_{[b\bar{b}]}
\left[\frac{d\hat{\sigma}_{ab\rightarrow[b\bar{b}](Q)}}
           {dQ^2dp_T^2 dy}\left(Q^2=M_{\Upsilon}^2\right)\right]
\int dQ^2\, {\cal F}_{[b\bar{b}]\rightarrow\Upsilon}(Q^2)
+{\cal O}\left(\frac{\langle q^2\rangle}{M_{\Upsilon}^2}\right)\, ,
\label{parton:xsec:NRQCD}
\end{equation}
with $m_b=M_{\Upsilon}/2$.  The integral 
$\int dQ^2\, {\cal F}_{[b\bar{b}]\rightarrow\Upsilon}(Q^2)
\equiv \langle \hat{O}_{[b\bar{b}]}(0)\rangle$
corresponds to a local matrix element of the $b\bar{b}$ pair 
in the NRQCD model. In the NRQCD approach to heavy quarkonium 
production, the $b \bar{b}$ pair need not have 
the quantum numbers of the $\Upsilon$.  It is assumed that 
non-perturbative soft gluons take care of the spin and color of 
the $\Upsilon$.  The sum in Eq.~(\ref{parton:xsec:NRQCD}) 
runs over all spin and color states of the $b \bar{b}$ system.

For the purpose of calculating the inclusive $p_T$ distributions of 
$S$-wave bound states at large enough $p_T$, both the CEM in
Eq.~(\ref{parton:xsec:CEM}) and the leading order NRQCD approach in
Eq.~(\ref{parton:xsec:NRQCD}) are expected to yield distributions
similar in shape because of the relatively weak $Q^2$ dependence of
the partonic hard-part in the limited range of $Q^2$.  For example,
$p_T$ distributions of $J/\psi$ and $\psi'$ production at
Tevatron energies are consistent with both CEM~\cite{Amundson:1996qr} 
and NRQCD~\cite{reviews} calculations for $p_T
\ge 5$~GeV.  
For production of $\Upsilon(nS)$
states, we choose a ${\cal F}(Q^2)$ that covers both the CEM and 
main properties of the leading order NRQCD treatment of heavy
quarkonium production~\cite{Qiu:1998rz} 
\begin{equation}
{\cal F}_{[b\bar{b}]\rightarrow\Upsilon(nS)}(Q^2)
= \left\{ \begin{array}{lll}
          C_{\Upsilon(nS)} (1-z)^{\alpha_{\Upsilon(nS)}}  
        & {\hskip 0.2in} 
        &  \mbox{if}\quad M_{\Upsilon(nS)}^2 \leq Q^2 \leq 4M_B^2 \\
          0             
          &    
          &\mbox{otherwise}
          \end{array}
  \right.
\label{F:QVZ}
\end{equation}
with $z=(Q^2-M_{\Upsilon(nS)}^2)/(4M_B^2-M_{\Upsilon(nS)}^2)$. 
In Eq.~(\ref{F:QVZ}), $C_{\Upsilon(nS)}$ and 
$\alpha_{\Upsilon(nS)}$ are parameters determined from data as 
discussed in Sec~\ref{sec:results}.    
With the choice of ${\cal F}(Q^2)$ in Eq.~(\ref{F:QVZ}), we reproduce 
the CEM by setting $\alpha_{\Upsilon(nS)}=0$ and replacing the lower limit  
$M_{\Upsilon(nS)}^2$ by $4m_b^2$.  Other than the color degree of freedom, 
we could mimic the features of NRQCD by 
choosing a very large value for $\alpha_{\Upsilon(nS)}$.

With our choice of ${\cal F}(Q^2)$, the transverse
momentum distribution of $\Upsilon$ production becomes 
\begin{equation}
\frac{d\sigma_{AB\rightarrow\Upsilon(nS) X}}{dp_T^2 dy}
= C_{\Upsilon(nS)} \int_{M_{\Upsilon(nS)}^2}^{4M_B^2} dQ^2 
\left[
      \frac{d\sigma_{AB\rightarrow b\bar{b}(Q) X}}{dQ^2dp_T^2 dy}
\right]
\left(1-\frac{Q^2-M_{\Upsilon(nS)}^2}{4M_B^2-M_{\Upsilon(nS)}^2}
\right)^{\alpha_{\Upsilon(nS)}}\, .
\label{hadron:xsec:BQW}
\end{equation}
The $b\bar{b}$ cross section is factored in terms of parton densities and 
the partonic cross section as
\begin{equation}
\frac{d\sigma_{AB\rightarrow b\bar{b}(Q) X}}{dQ^2dp_T^2 dy}
=\sum_{a,b} \int dx_a\, \phi_{a/A}(x_a)\, dx_b\, \phi_{b/B}(x_b)\, 
\frac{d\hat{\sigma}_{ab\rightarrow b\bar{b}(Q) X}}{dQ^2dp_T^2 dy}\, .
\label{hadron:QQ:xsec}
\end{equation}
The sum $\sum_{a,b}$ runs over gluon and light quark flavors up to 
and including charm.  The partonic cross sections,
$d\hat{\sigma}_{ab\rightarrow b\bar{b}(Q) X}/dQ^2dp_T^2 dy$ are 
computed at ${\cal O}(\alpha_s^3)$ from all $2$-parton to $3$-parton 
Feynman diagrams for the subprocesses $q\bar{q}\rightarrow b\bar{b}g$, 
$qg\rightarrow b\bar{b}q$, and $gg\rightarrow b\bar{b}g$, with 
the squared amplitudes summed over the spins and colors of the 
$b \bar{b}$ pair~\cite{Ellis:1986ef}.

\section{The region of small transverse momentum}
\label{sec:prodsmall}

When $p_T$ (or $Q_T$ of the $b\bar{b}$ pair) becomes small, the perturbatively 
calculated hard-part 
$d\hat{\sigma}_{ab\rightarrow [b\bar{b}](Q)X'}/dQ^2dp_T^2 dy$
in Eq.~(\ref{parton:xsec}) becomes singular
\begin{equation}
\frac{d\hat{\sigma}_{ab\rightarrow [b\bar{b}](Q)X'}}
           {dQ^2dp_T^2 dy}
\propto \frac{1}{p_T^2}.
\label{diverge}
\end{equation}
The $1/p_T^2$ singularity arises from the collinear region of
initial-state parton splitting.
Gluon radiation from the final-state heavy 
quark lines
does not contribute a $1/p_T^2$ 
collinear singularity because the heavy quark mass regulates this singularity.
However, this gluon radiation does lead to a $1/p_T^2$ infrared divergence 
which should be absorbed into the non-local transition probability
distribution  
${\cal F}(Q^2)$~\cite{QiuSterman:QQfac}.  
\begin{figure}[ht]
\centerline{\includegraphics[width=8cm]{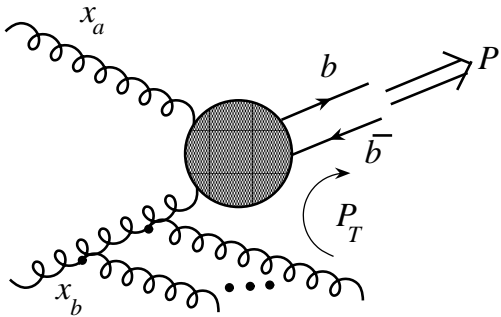}}
 \caption{Diagram that illustrates multiple gluon radiation 
 from an initial-state parton.} 
\label{fig3}
\end{figure}
When $p_T^2 \ll q^2=Q^2-4m_b^2$, soft gluon interactions between the 
spectator partons in the beam jets and the partons in ${\cal F}(Q^2)$ most 
likely break the factorization expressed in Eqs.~(\ref{hadron:xsec}) 
and~(\ref{parton:xsec}).  In this paper, our principal interest is to
investigate  
how the large logarithmic terms from the initial-state gluon shower modify
the $1/p_T^2$ distribution when $p_T^2 \ll M_{\Upsilon}^2$.

\subsection{Resummation of Sudakov logarithms in $b$-space}

Additional gluon radiation from the initial-state partons, recoiling against 
the $b\bar{b}$ pair as shown in Fig.~\ref{fig3}, leads to (Sudakov)
logarithmic  
contributions of the form $\alpha_s\log^2(Q^2/p_T^2)$ for each gluon 
radiation~\cite{Laenen:2000ij}.  The effects 
of the large Sudakov logarithmic contributions, very important in the 
region of small $p_T$, can be resummed to all orders in $\alpha_s$ when 
$p_T \ll Q$~\cite{Dokshitzer:hw}.  The resummation procedure tames the 
divergence seen in Eq.~(\ref{diverge}). Adopting the Collins, Soper, and 
Sterman (CSS) impact-parameter $b$-space (Fourier conjugate to $p_T$) 
approach~\cite{Collins:1984kg}, we write the resummed transverse momentum
distribution for $b\bar{b}$ production as 
\begin{eqnarray}
\frac{d\sigma_{AB\rightarrow [b\bar{b}](Q) X'}^{\rm resum}}
     {dQ^2dp_T^2dy}
&=&
\frac{1}{(2\pi)^2}\int d^2b\, e^{i\vec{p}_T\cdot \vec{b}}\,
{\cal W}_{AB\rightarrow [b\bar{b}](Q)}(b,Q,x_A,x_B)
\nonumber \\
&=& 
\int \frac{db}{2\pi}\, J_0(p_T\, b)\, 
b{\cal W}_{AB\rightarrow [b\bar{b}](Q)}(b,Q,x_A,x_B) \, .
\label{resum:QQ}
\end{eqnarray}
The function ${\cal W}_{AB\rightarrow [b\bar{b}](Q)}(b,Q,x_A,x_B)$ resums to 
all orders in QCD perturbation theory the singular terms from initial-state 
gluon showers that otherwise 
behave as $\delta^2(p_T)$ and $(1/p_T^2)\log^m(Q^2/p_T^2)$ for all $m \ge 0$.
In Eq.~(\ref{resum:QQ}), the fractional partonic momenta are 
$x_A= \frac{Q}{\sqrt{S}}\, e^y$ and 
$x_B= \frac{Q}{\sqrt{S}}\, e^{-y}$, with $\sqrt{S}$
the overall center-of-mass collision energy, and 
$y$ the rapidity of the $b\bar{b}$ pair; $x_A$ and $x_B$ 
are independent of the transverse momentum $p_T$ of the pair.  
The entire dependence on $p_T$ appears in the 
argument of the Bessel function $J_0$.  

The expressions for the lowest order subprocesses $gg\rightarrow b\bar{b}$ and 
$q\bar{q}\rightarrow b\bar{b}$ are independent of $p_T$.  Therefore,
the finite  
lowest order partonic cross sections can be used as prefactors in the overall 
$b$-space distribution functions~\cite{Laenen:2000ij}.
We write 
\begin{eqnarray}
{\cal W}_{AB\rightarrow [b\bar{b}](Q)}(b,Q,x_A,x_B) 
&\equiv & \sum_q W_{q\bar{q}}(b,Q,x_A,x_B) 
\frac{d\hat{\sigma}^{\rm (LO)}_{q\bar{q}\rightarrow [b\bar{b}](Q)}}
     {dQ^2}
\nonumber\\
&+& W_{gg}(b,Q,x_A,x_B) 
\frac{d\hat{\sigma}^{\rm (LO)}_{gg\rightarrow [b\bar{b}](Q)}} {dQ^2} \, .
\label{w:fac:sigma0}
\end{eqnarray}
The sum $\sum_q$ runs over all flavors of light quarks in the initial
state.  
The lowest order partonic cross sections in Eq.~(\ref{w:fac:sigma0}), 
$d\hat{\sigma}_{ij\rightarrow [b\bar{b}]}^{\rm (LO)}(Q)/dQ^2$ with 
$ij=q\bar{q},gg$, depend on the choice of the production model.  For
our choice of ${\cal F}_{b\bar{b}\rightarrow\Upsilon}(Q^2)$, they are  
\begin{eqnarray}
       \frac{d \hat{\sigma}^{\rm (LO)}_{ij\rightarrow b\bar{b}}(Q)}
            {dQ^2}
     =\sum_{[b\bar{b}]}
       \frac{d \hat{\sigma}^{\rm (LO)}_{ij\rightarrow[b\bar{b}]}(Q)}
            {dQ^2}
     = \frac{x_a x_b}{Q^2}\, \hat{\sigma}^{\rm (LO)}_{ij}(Q^2)
\end{eqnarray}
with~\cite{CEMearly}
\begin{eqnarray}
\hat{\sigma}_{q\bar{q}}^{\rm (LO)}(Q^2) 
&=&\frac{2}{9}\, \frac{4\pi\alpha_s^2}{3Q^2}
   \left[ 1 + \frac{1}{2}\gamma\right]\sqrt{1-\gamma} \, ;
\nonumber \\
\hat{\sigma}_{gg}^{\rm (LO)}(Q^2) 
&=&\frac{\pi\alpha_s^2}{3Q^2}
   \left[\left(1+\gamma+\frac{1}{16}\gamma^2\right)
          \ln\left(\frac{1+\sqrt{1-\gamma}}{1-\sqrt{1-\gamma}}\right)
        -\left(\frac{7}{4}+\frac{31}{16}\gamma\right)\sqrt{1-\gamma}
   \right] \, ;
\label{QQ:xsec:lo}
\end{eqnarray}
and $\gamma=4m_b^2/Q^2$.

When the impact parameter $b$ lies in the region much less than 
1 GeV$^{-1}$ where perturbation theory applies, the distributions 
$W_{q\bar{q}}(b,Q,x_A,x_B)$ and $W_{gg}(b,Q,x_A,x_B)$ in 
Eq.~(\ref{w:fac:sigma0}) can be expressed as~\cite{Collins:1984kg}
\begin{equation}
W_{ij}^{\rm pert}\left(b,Q,x_A,x_B\right) 
= {\rm e}^{-S_{ij}(b,Q)}\, 
  f_{i/A}\left(x_A,\mu,\frac{c}{b}\right)\, 
  f_{j/B}\left(x_B,\mu,\frac{c}{b}\right)\,
  H_{ij}\, ,
\label{cfg-W-H}
\end{equation}
where $ij=q\bar{q}$ and $gg$, $\mu$ is the factorization scale,
and $c=2e^{-\gamma_E}={\cal O}(1)$, with Euler's constant
$\gamma_E\approx 0.577$.  
All large Sudakov logarithmic terms from 
$\log(c^2/b^2)$ to $\log(Q^2)$ 
are resummed to all orders in $\alpha_s$ in the exponential 
factors with  
\begin{equation}
S_{ij}(b,Q) = \int_{c^2/b^2}^{Q^2}\, 
  \frac{d\bar{\mu}^2}{\bar{\mu}^2} \left[
  \ln\left(\frac{Q^2}{\bar{\mu}^2}\right) 
     {\cal A}_{ij}(\alpha_s(\bar{\mu})) 
   + {\cal B}_{ij}(\alpha_s(\bar{\mu})) \right]\, .
\label{css-S}
\end{equation}
The functions ${\cal A}_{ij}$ and ${\cal B}_{ij}$ may be 
expanded in perturbative power series in  
$\alpha_s$; ${\cal A}_{ij}=\sum_{n=1} {\cal A}_{ij}^{(n)}
\left(\alpha_s/\pi\right)^n$, and 
${\cal B}_{ij}=\sum_{n=1} {\cal B}_{ij}^{(n)}
\left(\alpha_s/\pi\right)^n$.  
The first two coefficients in the power series for ${\cal A}_{ij}$ 
and the first term in the series for ${\cal B}_{ij}$ are 
{\it process-independent}.  For $ij=q\bar{q}$  
they are the same as those that appear in resummation of the 
transverse momentum distribution for massive-lepton-pair production 
(Drell-Yan production)~\cite{Collins:1984kg,Davies:1984sp,Qiu:2000hf}.
For $ij=gg$, they are the same  
as the coefficients that are appropriate for resummation of the 
$p_T$ distribution of Higgs boson 
production~\cite{Catani:vd,Catani:2000vq,Berger:2002ut,Kulesza:2003wi}.  

The modified parton distribution functions in Eq.~(\ref{cfg-W-H}) are
expressed as~\cite{Collins:1984kg}
\begin{equation}
f_{i/A}\left(x_A,\mu,\frac{c}{b}\right) = \sum_a 
  \int_{x_A}^1\frac{d\xi}{\xi}\, \phi_{a/A}(\xi,\mu)\,
  {\cal C}_{a\rightarrow i}
       \left(\frac{x_A}{\xi},\mu,\frac{c}{b}\right)\, .
\label{mod-pdf}
\end{equation} 
In Eq.~(\ref{mod-pdf}), $\phi_{a/A}$ are the usual parton distribution   
functions.  The functions 
${\cal C}_{a\rightarrow i}=\sum_{n=0} {\cal C}_{a\rightarrow
  i}^{(n)}\left(\alpha_s/\pi\right)^n$ are $b$-space 
coefficient functions with the lowest order terms normalized to 
${\cal C}^{(0)}_{a\rightarrow i}(z,\mu,c/b) = 
\delta_{ai}\,\delta(1-z)$.
All higher order coefficient functions are computed perturbatively
from the Fourier transform of the singular terms in $p_T$-space from
initial-state gluon showers 
with $Q^2=c^2/b^2$. 

In the CSS formalism, $H_{ij}=1$ in Eq.~(\ref{cfg-W-H}).   
All coefficient functions, ${\cal C}^{(n)}_{a\rightarrow i}$ with
$n\ge 1$, are {\it process-dependent} representing the 
non-logarithmic short-distance partonic contributions to 
$\sigma^{\rm resum}$.
Alternatively, one may be able to reorganize Eq.~(\ref{cfg-W-H}) 
such that all {\it process-dependent} short-distance contributions are
moved into the {\it process-dependent} hard part $H_{ij}$, leaving 
the coefficient functions ${\cal C}_{a\rightarrow i}$ and the 
modified parton distributions 
{\it process-independent}~\cite{Catani:2000vq}.
Expressions for $H_{ij}^{n}$ with
$n\ge 1$ depend on the ``resummation scheme'', the {\it choices} 
made when the {\it process-dependent} finite pieces are moved 
from the higher order terms in the ${\cal A}$'s, ${\cal B}$'s, and 
${\cal C}$'s to the $H_{ij}$ functions~\cite{Catani:2000vq}.
  
For production of a colorless object, such as the $W$, $Z$, and 
Higgs bosons or a virtual photon in the Drell-Yan process, all resummed 
 $(1/p_T^2)\log^m(Q^2/p_T^2)$ singular terms arise from initial-state
 gluon showers.  For the $\Upsilon$, which is not
 produced directly in the hard collision, additional singular $1/p_T^2$ 
 terms can originate from soft gluon
 radiation from the $b\bar{b}$ pair.  These additional $1/p_T^2$ terms
 should not be included in the resummation of Sudakov logarithms from
 initial-state gluon showers.  The calculation of the $n\ge 1$ corrections 
 to the coefficient functions ${\cal C}^{(n)}_{a\rightarrow i}$ and hard 
 part $H^{(n)}_{ij}$ (or ${\cal C}^{(n)}_{a\rightarrow i}$ in the CSS 
 formalism) should involve a systematic removal of these additional 
 $1/p_T^2$ singular terms.

The long-distance nature of soft-gluon radiation means that the additional 
singular terms from final-state radiation should be included in the 
non-perturbative transformation of the $b\bar{b}$ pair to the $\Upsilon$.  
Therefore, the removal of the additional $1/p_T^2$ terms 
depends on the models of $\Upsilon$ production.  In terms of the
two-step factorization procedure discussed in Sec.~\ref{sec:prod}, 
the $1/p_T^2$ terms should be absorbed into the transition
probability distributions ${\cal F}_{[b\bar{b}]\rightarrow 
  \Upsilon X}$ defined in terms of matrix elements of non-local
operators.  In the NRQCD model of heavy quarkonium production, 
these $1/p_T^2$ singularities are a consequence of the kinematic 
end-point of the quarkonium transverse momentum spectrum. 
Although the kinematic effect of soft-gluon emission from the heavy
quark pair is usually a higher-order effect in the non-relativistic
expansion, the high order non-perturbative contributions are enhanced in
the region of the kinematic end point as $p_T\rightarrow 0$, leading to a
breakdown of the NRQCD expansion and the introduction of ``shape
functions''~\cite{Beneke:1997qw,Beneke:1999gq}. 

Faced with the model dependence of the $\alpha_s$ corrections
and the complication of separating singular terms of different
origins, we resum only {\it process-independent} logarithmic terms
from initial-state gluon showers in this paper.   
That is, we keep only ${\cal A}_{q,g}^{(1)}$, ${\cal A}_{q,g}^{(2)}$, 
and ${\cal  B}_{q,g}^{(1)}$ in the Sudakov exponential functions 
$S_{q,g}(b,Q)$ in Eq.~(\ref{css-S}), the lowest order coefficient
function ${\cal C}_{a\rightarrow i}^{(0)}$, and the lowest order
short-distance hard parts $H^{(0)}_{ij}=1$ in Eq.~(\ref{cfg-W-H}).
We choose the factorization scale $\mu=c/b$ for the resummed 
$b$-space distribution in Eq.~(\ref{cfg-W-H})~{\footnote{Since 
there is only one momentum scale, $1/b$, involved in the  
functions $W_{ij}(b,c/b,x_A,x_B)$, it is natural to choose 
$\mu=c/b$ in the modified parton densities $f(x,\mu,c/b)$ in order 
to remove the logarithmic terms in the coefficient functions, 
${\cal C}$.  However, this choice is not required.  For more 
discussion, see Ref.~\cite{Berger:2002ut}.}}. We defer to a 
future study the calculation of {\it  process- and model-dependent}
higher order corrections ${\cal C}^{(1)}_{a\rightarrow i}$ and 
$H^{(1)}_{ij}$ as well as ${\cal A}^{(3)}$ and ${\cal B}^{(2)}$.  
As a result of these restrictions, we anticipate that our calculation
will somewhat underestimate the magnitude of the differential cross
section in the region of small $p_T$, and we return to this point in
next section. 
 
\subsection{Predictive power}

The predictive power of the Fourier transformed formalism in 
Eq.~(\ref{resum:QQ}) depends critically on the shape of the $b$-space 
distribution function $b{\cal W}(b,Q,x_A,x_B)$~\cite{Qiu:2000hf}.  
Indeed, the resummed calculation of the transverse momentum distribution 
at low $p_T$ can be reliable {\it only if} the Fourier transformation 
in Eq.~(\ref{resum:QQ}) is dominated by the region of small $b$, where 
perturbation theory applies, and is not sensitive to the extrapolation
to the region of large $b$.  
This condition is achieved if the 
distribution $b{\cal W}(b,Q,x_A,x_B)$ has a prominent saddle point for  
$b_{\rm sp}\ll 1$~GeV$^{-1}$.

The location of the saddle point in the $b$-space distribution depends 
not only on the value of $Q$ but also strongly on the collision energy
$\sqrt{S}$ (or, equivalently, on the values of the parton momentum 
fractions $x_a$ and $x_b$ that control the cross section)~\cite{Qiu:2000hf}.  
At $\sqrt{S}=1.8$~TeV, the saddle point can be within the perturbative region 
($b_{\rm sp}< 0.5$~GeV$^{-1}$) for $Q$ as low as 6~GeV~\cite{Qiu:2000hf}.  

An extrapolation into the region of large 
$b$ is needed in order for us to perform the Fourier transformation to 
the $p_T$ distribution in Eq.~(\ref{resum:QQ}).  We choose the Qiu-Zhang 
prescription which has the desirable property that it separates cleanly 
the perturbative prediction at small $b$ from non-perturbative contributions 
in the large $b$ region.  
\begin{equation}
W_{ij}(b,Q,x_A,x_B) = \left\{
\begin{array}{ll}
 W^{\rm pert}_{ij}(b,Q,x_A,x_B) &  \mbox{$b\leq b_{max}$} \\
 W^{\rm pert}_{ij}(b_{max},Q,x_A,x_B)\,
                    F^{NP}_{ij}(b,Q;b_{max})
                                       &  \mbox{$b > b_{max}$}
\end{array} \right.
\label{qz-W-b}
\end{equation}
for $ij=q\bar{q}$ and $gg$.  
The perturbative distribution $W^{\rm pert}_{ij}(b,Q,x_A,x_B)$
is given in Eq.~(\ref{cfg-W-H}).
The nonperturbative function in the 
large $b$ region is 
\begin{eqnarray}
F^{NP}_{ij}
&=&
\exp\bigg\{
 -\ln(\frac{Q^2 b_{max}^2}{c^2}) \left[
   g_1 \left( (b^2)^\alpha - (b_{max}^2)^\alpha\right)
  +g_2 \left( b^2 - b_{max}^2\right) \right]
\nonumber \\
&\ & \hskip 0.4in
 -\bar{g}_2 \left( b^2 - b_{max}^2\right)
  \bigg\} .
\label{qz-fnp}
\end{eqnarray}
The $(b^2)^\alpha$ term with $\alpha<1/2$ represents a direct
extrapolation of the resummed function $W^{\rm pert}_{ij}(b,Q,x_A,x_B)$.
The parameters, $g_1$ and $\alpha$ are fixed from  
$W^{\rm pert}_{ij}$ if we require that the first and second
derivatives of $W_{ij}(b,Q,x_A,x_B)$ be continuous at $b=b_{max}$.   
The $x_A$ and $x_B$ dependences of the nonperturbative function
$F^{NP}_{ij}$ are included in the parameters $g_1$ and $\alpha$.

The two terms proportional to $b^2$ correspond to power corrections in 
the evolution equation.  The $g_2$ term represents a power correction 
from soft gluon showers.   The $\bar{g}_2$ term is associated with the 
finite intrinsic transverse momentum of the incident partons.  
Since $g_1$ and $\alpha$ are fixed by the continuity of the 
$W_{ij}(b,Q,x_A,x_B)$ at $b=b_{max}$, the (in)sensitivity of the calculated
$p_T$ distribution to the numerical values of $b_{max}$, $g_2$, and
$\bar{g}_2$ is a good quantitative measure of the predictive power of
the resummmation formalism~\cite{Qiu:2000hf}.

\begin{figure}[ht]
\centerline{\includegraphics[width=14cm]{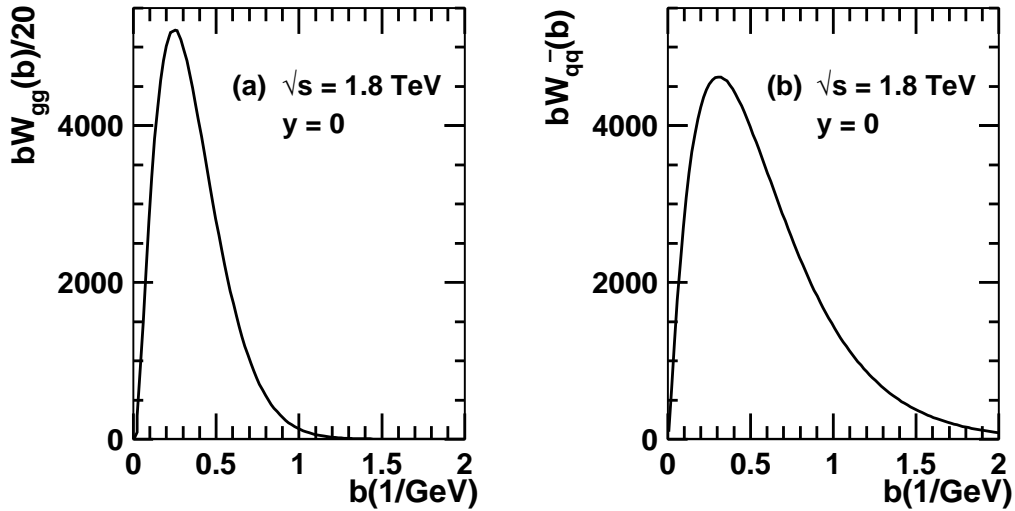}}
\caption{The $b$-space distributions for $\Upsilon$ production:
(a) $gg$ channel, and (b) {\em the sum of all} $q\bar{q}$ channels. Note 
that the magnitude of the $gg$ distribution has been scaled by a factor of 20. 
The functions are evaluated at rapidity $y = 0$ and are integrated 
over the mass range $2m_b < Q < 2M_B$.}
\label{fig4}
\end{figure}

In Fig.~\ref{fig4}, we show the $b$-space distributions that result from 
Eq.~(\ref{qz-W-b}) at rapidity {$y = 0$}.  The functions are integrated  
over the mass range $2m_b < Q < 2M_B$.  In evaluating the
perturbative distribution, we keep only the {\it process-independent}
terms in the Sudakov exponential functions $S_{q,g}(b,Q)$ in
Eq.~(\ref{css-S}) and use the CTEQ6M parton
densities~\cite{Pumplin:2002vw}.   
For the extrapolation to the large $b$ region, we choose
$b_{max}=0.5$~GeV$^{-1}$ and $g_2=\bar{g}_2=0$.
The magnitude of $bW_{gg}(b,Q,x_A,x_B)$ is 
scaled by a factor of 20, as is indicated in the figure.  The 
$gg$ contribution far exceeds the $q \bar{q}$ contribution to $\Upsilon$ 
production at Tevatron energies.  
For both channels, the saddle points are clearly defined and 
have numerical values well within the perturbative region.  For the dominant 
$gg$ channel, the location of the saddle point is 
$b_{\rm sp} \sim 0.25$~GeV$^{-1}$.  For perspective, we remark 
that this value is smaller than that for the saddle point of $W$ and $Z$ boson 
production at Tevatron energies~\cite{Qiu:2000hf}.  This feature arises 
because the gluon anomalous dimension at small $x$ is much larger than that of 
the quarks, compensating for the fact that the mass $Q$ here is much less than 
the mass of $W$ and $Z$ bosons. 
This analysis leads us to expect that the QCD resummed $p_T$ distribution of
$\Upsilon$ production in Eq.~(\ref{resum:QQ}) can be predicted reliably in 
the region of small and intermediate $p_T$    
because it is dominated by perturbative contributions in the region of 
small $b$.

\section{Numerical results}
\label{sec:results}

In this section 
we present the results of our numerical computation, including a 
comparison with data.  
We compute the $\Upsilon$ transverse momentum distribution from 
Eq.~(\ref{hadron:xsec:BQW}).  For the region of large transverse
momentum, $p_T \sim {\cal O} (M_{\Upsilon})$, 
the $b\bar{b}$ cross section is given in Eq.~(\ref{hadron:QQ:xsec}).
For the region of small $p_T$, we use the all-orders resummed
$b\bar{b}$ cross section in Eq.~(\ref{resum:QQ}) with the $b$-space
distribution ${\cal W}_{AB\rightarrow b\bar{b}}$ specified in
Eqs.~(\ref{w:fac:sigma0}) and (\ref{qz-W-b}).
We set $m_b = 4.5$~GeV, and we use a two-loop 
expression for $\alpha_s$, in keeping with our use of the CTEQ6M 
parton densities~\cite{Pumplin:2002vw}. 

To distinguish the production of $\Upsilon(nS)$ states with different
$n$, we choose different powers $\alpha_{\Upsilon(nS)}$ and normalization 
constants $C_{\Upsilon(nS)}$,   
in addition to the differences in mass threshold on the limits of the 
$dQ^2$ integration in Eq.~(\ref{hadron:xsec:BQW}).  The values of 
$\alpha_{\Upsilon(nS)}$ and $C_{\Upsilon(nS)}$ are correlated.  A larger 
value of $\alpha_{\Upsilon(nS)}$ leads to a larger value of
$C_{\Upsilon(nS)}$.

\subsection {Matching of results at small and large $p_T$}

In a complete calculation, one would expect a seamless joining of the
results applicable at small and at large $p_T$.  In the CSS
resummation formalism for production of a color singlet heavy boson,
this matching is accomplished through  
the introduction of an ``asymptotic'' term, $\sigma^{\rm asym}$, and  
\begin{equation}
d\sigma = d\sigma^{\rm resum} + (d\sigma^{\rm pert} 
        - d\sigma^{\rm asym}) \, .
\end{equation}
The term $\sigma^{\rm asym}$ is constructed to cancel the singular
behavior of $\sigma^{\rm pert}$ as $p_T \rightarrow 0$ and to cancel
$\sigma^{\rm resum}$ when $p_T \sim Q$.  It is obtained from the
fixed-order terms in the expansion of  
$\sigma^{\rm resum}$ in a power series in $\alpha_s$.  
 
The procedure just described is not immediately applicable in our
case.  Because the $b\bar{b}$ system is not necessarily in a color
neutral state, $\sigma^{\rm pert}$ includes radiation from the heavy
quark system as well as from the incoming partons. 
This final state radiation is not included in either 
$\sigma^{\rm resum}$ or $\sigma^{\rm asym}$.  
Soft gluon radiation from the heavy 
quark system leads to an infrared divergent $1/p_T^2$
singularity~\cite{Berger:yp} which should be absorbed into 
${\cal F}(Q^2)$~\cite{QiuSterman:QQfac}. 
To avoid extrapolation of $\sigma^{\rm pert}$ into region of low
$p_T$, we adopt the following matching procedure:
\begin{equation}
\frac{d\sigma_{AB\rightarrow \Upsilon(nS)X}}{dp_T^2 dy}
= \left\{ \begin{array}{lll}
      \frac{d\sigma_{AB\rightarrow \Upsilon(nS)X}^{\rm resum}}
               {dp_T^2 dy}
                        & {\hskip 0.2in} 
                        &  p_T < p_{T_M} \\
      \frac{d\sigma_{AB\rightarrow \Upsilon(nS)X}^{\rm pert}}
               {dp_T^2 dy}
                        &    
                        &  p_T \ge p_{T_M} .
          \end{array}
  \right.
\label{match:pt}
\end{equation} 
Matching is done at a value $p_{T_M}$  chosen as the 
location of intersection of the resummed and perturbative components of 
the $p_T$ distribution.  From other work on resummed $p_T$ 
spectra~\cite{Davies:1984sp,Qiu:2000hf,Balazs:1997hv,Berger:yp,Catani:vd,Catani:2000vq,Berger:2002ut},
we expect $p_{T_M} \sim M_{\Upsilon}/2$.  

To ensure a smooth parameter-free matching of $\sigma^{\rm resum}$ to 
the perturbative $p_T$ distribution $\sigma^{\rm pert}$ computed at 
${\cal O}(\alpha_s^3)$, we would need to calculate the {\it process-dependent} 
${\cal O}(\alpha_s)$ corrections ${\cal C}^{(1)}_{a\rightarrow i}$ 
and $H^{(1)}_{ij}$ (or ${\cal C}^{(1)}_{a\rightarrow i}$ in the CSS 
formalism) for $\sigma^{\rm resum}$.  If these ${\cal O}(\alpha_s)$ corrections
were included, $\sigma^{\rm resum}$ would also be of order ${\cal O}(\alpha_s^3)$
at the matching point where the logarithms are not important.
Based on prior experience~\cite{Berger:2002ut}, we expect that these
effects will increase the predicted 
normalization of $d\sigma^{\rm resum}/dp_T^2 dy$, and change the shape of 
the $p_T$ distribution somewhat, increasing (decreasing) the spectrum at 
small (large) $p_T$.  For the reasons stated in last section, we do
not calculate the order $\alpha_s$ corrections to 
$\sigma^{\rm resum}$ in this paper.  
To account for the size of the order $\alpha_s$ corrections, we 
introduce a {\em resummation enhancement factor} $K_r$ such that
\begin{equation}
\left[{\cal C}^{(0)}_{a\rightarrow i}
     + {\cal C}^{(1)}_{a\rightarrow i}\, \frac{\alpha_s}{\pi} \right]
\otimes
\left[{\cal C}^{(0)}_{b\rightarrow j}
    + {\cal C}^{(1)}_{b\rightarrow j}\, \frac{\alpha_s}{\pi} \right]
\otimes
\left[H_{ij}^{(0)} + H_{ij}^{(1)}\, \frac{\alpha_s}{\pi} \right]
\equiv \, K_r \,
{\cal C}^{(0)}_{a\rightarrow i}
\otimes
{\cal C}^{(0)}_{a\rightarrow i}
\otimes
H_{ij}^{(0)} \, .
\label{Kr}
\end{equation}
We assume that $K_r$ is a constant.
The factor $K_r$ should not be confused with a ``$K$-factor'' for
the overall $p_T$ distribution.  It is invoked because we do not
calculate the order $\alpha_s$ corrections to $\sigma^{\rm resum}$, 
and our $\sigma^{\rm resum} \sim {\cal O}(\alpha_s^2)$ when
$p_T\sim {\cal O}(M_\Upsilon)$.

\begin{figure}[ht]
\centerline{\includegraphics[width=12cm]{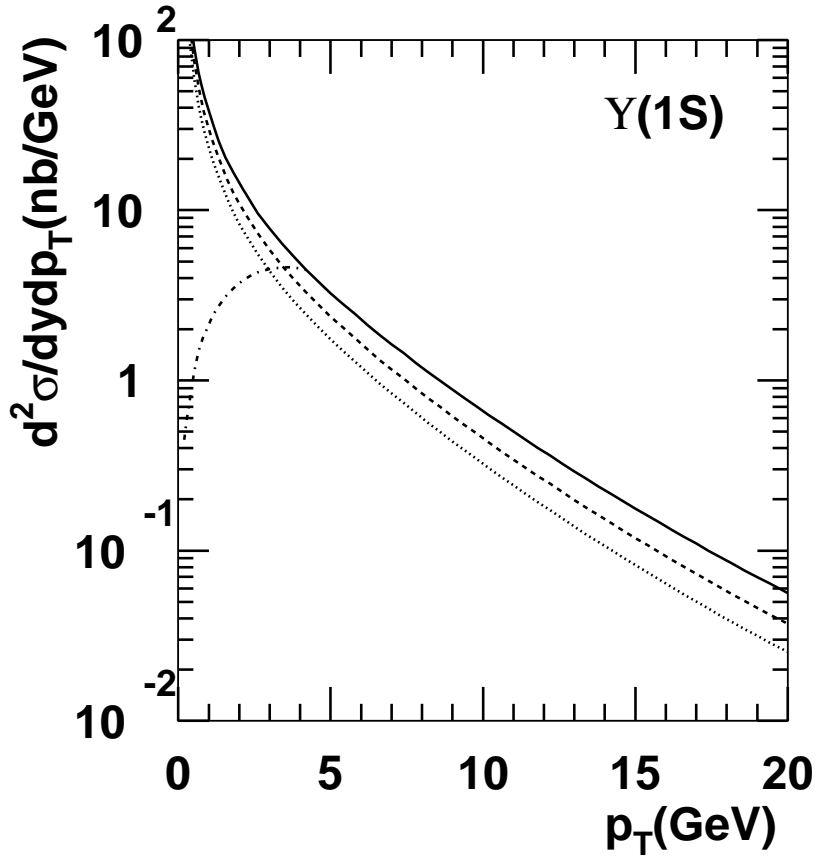}}
\caption{Inclusive transverse momentum distribution of the 
$\Upsilon(1S)$ in $p \bar{p}$ interactions at $\sqrt S = 1.8$~TeV. 
The solid, dashed, and dotted lines in the region of large $p_T$ are 
obtained from fixed-order ${\cal O} (\alpha_s^3)$ perturbative 
QCD for three different values of the scale $\mu$; solid for 
$\mu = 0.5\mu_0$, dashed for $\mu = \mu_0$, and dotted for 
$\mu = 2\mu_0$; with $\mu_0 = \sqrt{Q^2+p_T^2}$.  The dot-dashed 
line in the region of small $p_T$ is the resummed prediction.}
\label{fig5}
\end{figure}

Displayed in Fig.~\ref{fig5} are curves that show the differential  
cross section for production of the $\Upsilon(1S)$ as a function of
$p_T$.  The curves in the region of large $p_T$ illustrate the
dependence of the fixed-order ${\cal O} (\alpha_s^3)$ perturbative
cross section on the common renormalization/factorization scale $\mu$.
We vary $\mu$ over the range $0.5 < \mu/\mu_0 < 2$ where  
$\mu_0 = \sqrt{Q^2+p_T^2}$.  This variation demonstrates the
inevitable theoretical uncertainty of a fixed-order calculation. It
could be reduced if a formidable ${\cal O} (\alpha_s^4)$ calculation 
is done in perturbation theory.  We fix $\mu = 0.5 \sqrt{Q^2+p_T^2}$ 
for the remainder of our discussion.  The $1/p_T^2$ divergence mentioned 
in Eq.(\ref{diverge}) is evident in the fixed-order curves. Shown for
purposes of perspective is a dot-dashed line that represents the
resummed prediction, applicable at small $p_T$, obtained with
$b_{max}=0.5$~GeV$^{-1}$ and $g_2=\bar{g}_2=0$.  
The set of curves illustrates the range of possibilities 
for the value of the matching point $p_{T_M}$.   
To obtain these results, we set $C_{\Upsilon} = 0.044$,
$\alpha_{\Upsilon} = 0$, and $K_r = 1.22$, 
for reasons that are explained in the next subsection. 

%

\subsection{Comparison with data}

In order to make contact with data we must determine values for $K_r$ 
in Eq.~(\ref{Kr}) and $p_{T_M}$ in Eq.~(\ref{match:pt}), 
and for the two non-perturbative parameters $C_{\Upsilon}$  
and $\alpha_{\Upsilon}$ in the transition probability distribution 
${\cal F}(Q^2)$.  The structure of Eq.~(\ref{match:pt}) indicates that  
the data at {\it high} $p_T$ determine $C_{\Upsilon}$ and 
$\alpha_{\Upsilon}$, and the data at {\it low} $p_T$ fix the 
enhancement factor $K_r$. Dependence on the parameter 
$\alpha_{\Upsilon}$ turns out to be very weak, as might be expected
from the limited range in $Q^2$ over which ${\cal F}(Q^2)$ is
probed. This weak dependence confirms that the production models we
consider predict very similar inclusive $p_T$ distributions.  We
choose to set $\alpha_{\Upsilon}=0$ for all three $\Upsilon$  
states.  Second, common values of $K_r$ and $p_{T_M}$ work adequately
for all three $\Upsilon$ states, as might be expected since the
differences are small among the three Upsilon masses.

In our approach to the data, $C_{\Upsilon}$ represents not just 
the normalization in the transition probability distribution 
${\cal F}(Q^2)$, but the product of this normalization times the
unknown $K$-factor from order ${\cal O}(\alpha_s^4)$ perturbative
contributions at large $p_T$.  It should not be surprising that
non-perturbative free parameters enter the comparison with data at
large $p_T$.  The reliance on non-perturbative parameters to set the
normalization is true of all models other than the color-singlet  
model~\cite{reviews}.

We compare our calculation with data published by the Collider
Detector at Fermilab (CDF) collaboration~\cite{cdf2,Acosta:2001gv}
obtained in run I of the Tevatron collider at $\sqrt S = 1.8$~TeV.  In
the second of the two publications, it is noted that the measured
rates are lower than those reported in the first paper.  To account
for the difference in our fits to the data, we include an overall
multiplicative normalization factor $C_n$, whose value we determine
from our $\chi^2$ fitting routine.  This factor is used only for the
1995 data~\cite{cdf2}.   

Following our initial qualitative exploration of the data, we are left
with the three parameters $C_{\Upsilon}$, the common values of 
$K_r$ and $p_{T_M}$, and the data adjustment factor $C_n$.  We use a 
$\chi^2$ minimization procedure to determine these quantities.  We find best 
fit values $K_r = 1.22 \pm 0.02$, and $p_{T_M} \sim 4.27$~GeV.  The value of 
$K_r$ is comparable to typical $K$-factors found in next-to-leading order 
calculations, but, as remarked above, the origin here is different.  
The matching point is fixed essentially by the location where the 
resummed and perturbative cross section intersect.  Its value, 
$p_{T_M} \sim M_{\Upsilon}/2$, is similar to results 
found in other work on resummed $p_T$ 
spectra~\cite{Davies:1984sp,Qiu:2000hf,Balazs:1997hv,Berger:yp,Catani:vd,Catani:2000vq,Berger:2002ut}. 
We find that the values of $C_{\Upsilon}$:
$0.044 \pm 0.001$, $0.040 \pm 0.006$, and $0.041 \pm 0.003$ for 
$\Upsilon(1S)$, $\Upsilon(2S)$, and $\Upsilon(3S)$, respectively,
are approximately independent of $M_{\Upsilon}$, meaning that the  
differences in rates for the three $S$-wave $\Upsilon$ states are accounted 
for by the different threshold values of the integrals in
Eq.~(\ref{hadron:xsec:BQW}).  

In our determination of $C_\Upsilon$, $K_r$ and $p_{T_M}$, we
keep only {\it process-independent} terms in the Sudakov exponential
functions $S_{ij}(b,Q)$ in Eq.~(\ref{css-S}), and 
the parameters of the non-perturbative function $F^{NP}$ are 
fixed at $b_{max}=0.5$~GeV$^{-1}$, and $g_2=\bar{g}_2=0$.  Because 
of the dominance of the perturbative small-$b$ region under the curves of
$bW_{ij}$ in Fig.~\ref{fig4}, any reasonable values of $g_2$ and 
$\bar{g}_2$ lead to transverse momentum distributions that do not 
differ more than one percent from those calculated with 
$g_2=\bar{g}_2=0$~\cite{Qiu:2000hf,Berger:2002ut}.  Without adjusting
the normalization, we find a few percent change in the resummed
distributions over the entire low $p_T$ region when we vary $b_{max}$ 
from 0.3 to 0.7~GeV$^{-1}$.

\begin{figure}[ht]
\centerline{\includegraphics[width=5.4cm]{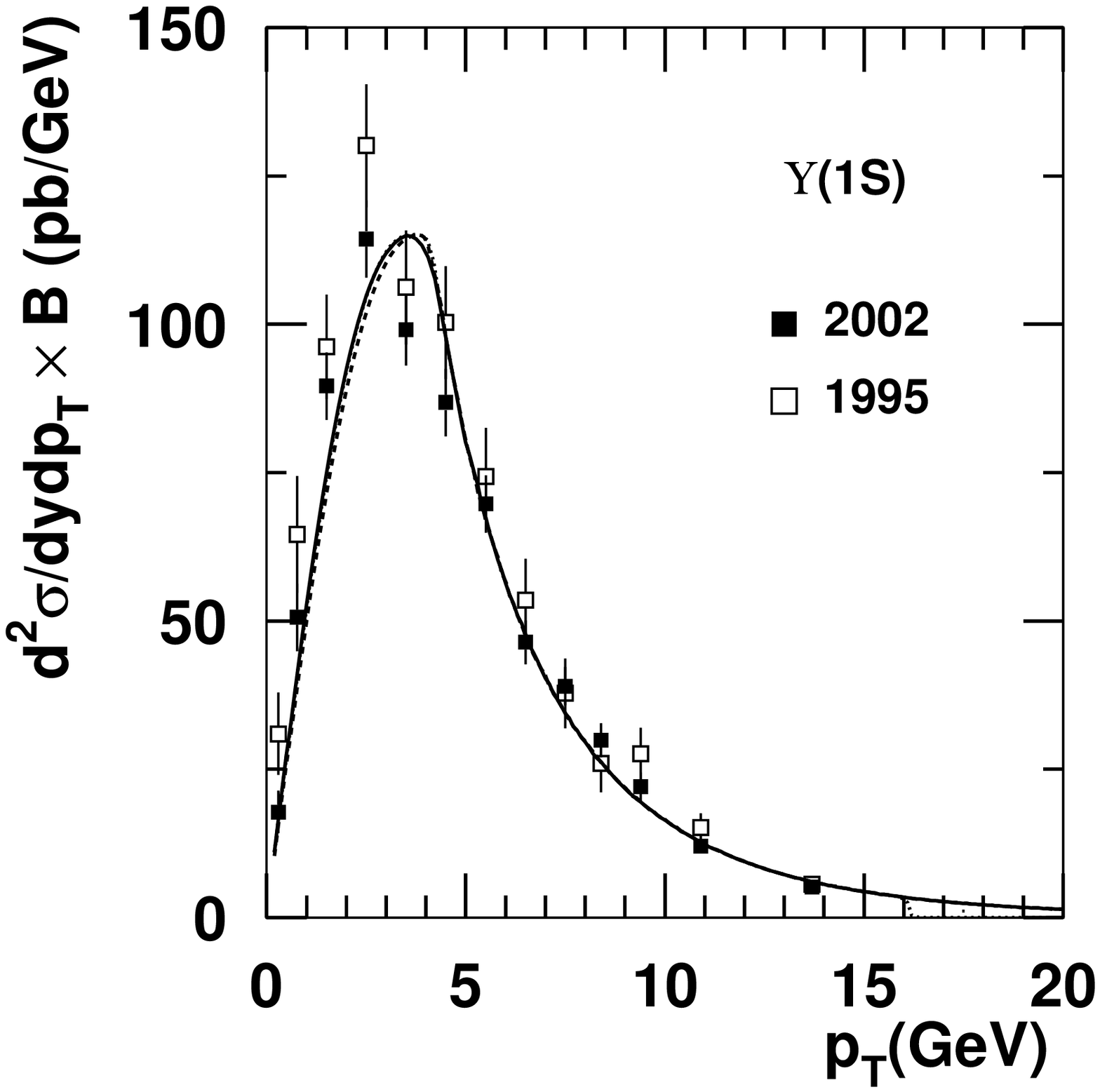}
\hfil       \includegraphics[width=5.4cm]{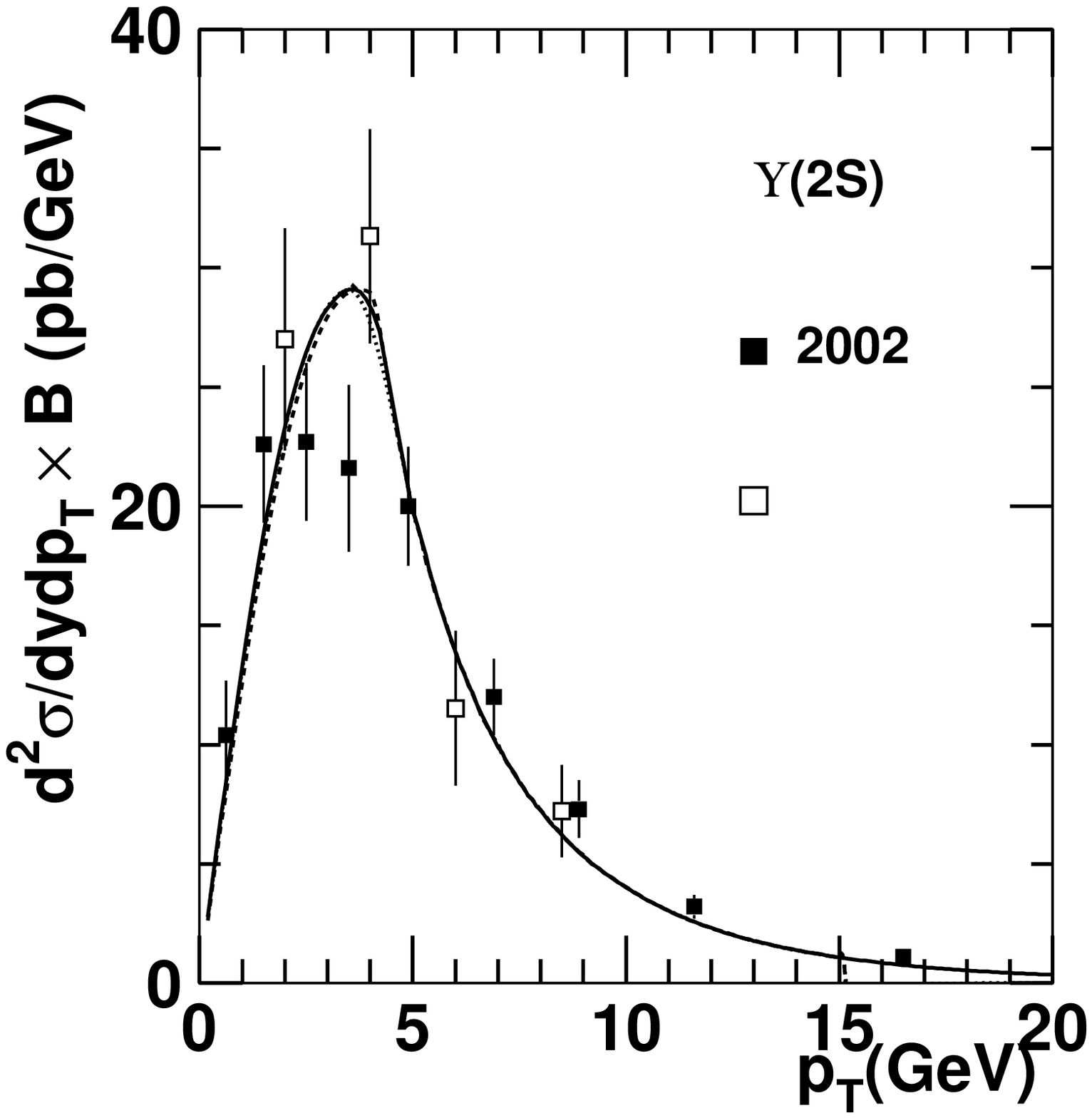}
\hfil       \includegraphics[width=5.4cm]{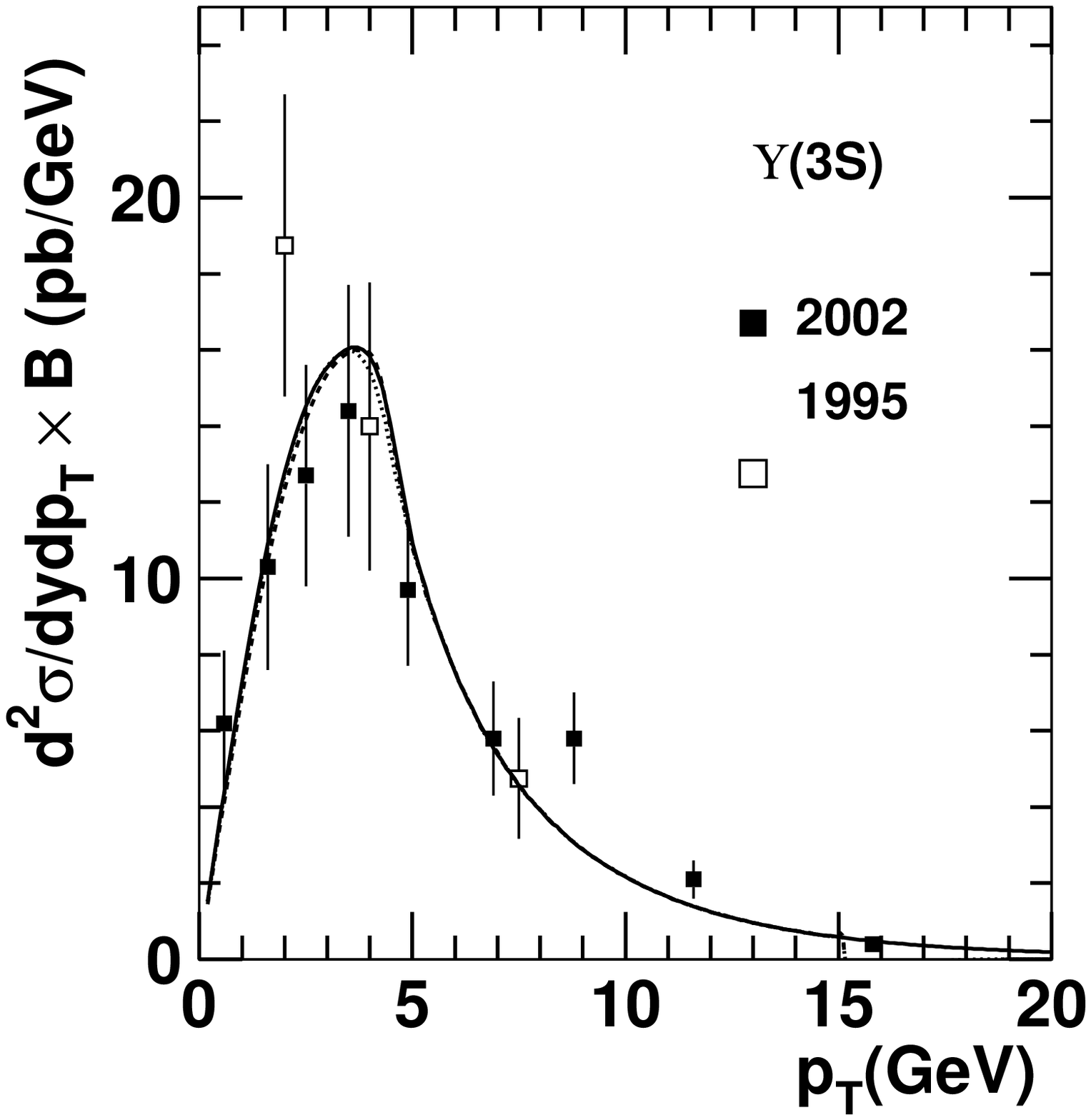}}
\caption{Calculated differential cross sections times leptonic branching 
fractions $B$, evaluated at $y = 0$, as functions of transverse momentum 
for hadronic production of 
(a) $\Upsilon(1S)$, (b) $\Upsilon(2S)$, and (c) $\Upsilon(3S)$, along with 
CDF data~\cite{cdf2,Acosta:2001gv} at $\sqrt S = 1.8$~TeV.  The dashed, 
solid, and dotted lines show the result of our full calculation
for $b_{max}=0.3$, 0.5, and 0.7~GeV$^{-1}$, respectively. 
The 1995 CDF cross sections are 
multiplied by $C_n = 0.88$.}
\label{fig6}
\end{figure}

The principal predictive power of our calculation is the shape of the 
$p_T$-distribution for the full $p_T$ region.   
In Fig.~\ref{fig6}, we present our calculation of the transverse  
momentum distribution for hadronic production of $\Upsilon(nS)$, $n =
1 - 3$, as  
obtained from our Eq.~(\ref{match:pt}), and multiplied by the leptonic  
branching fractions $B$.  We use the values of $B$ from
Ref.~\cite{Hagiwara:fs}.  
The solid lines are for $b_{max}=0.5$~GeV$^{-1}$ while the dashed and
dotted lines are for $b_{max}=0.3$ and 0.7~GeV$^{-1}$, respectively.
Also shown in Fig.~\ref{fig6} are data from the CDF 
collaboration~\cite{cdf2,Acosta:2001gv}.  We determine a data normalization
adjustment of $C_n = 0.88 \pm 0.05$ and use this value to multiply 
only the 1995 cross sections  
shown in the figure.  The shapes of the $p_T$ distributions 
are consistent with experimental results. 

\begin{figure}[ht]
\centerline{\includegraphics[width=14cm]{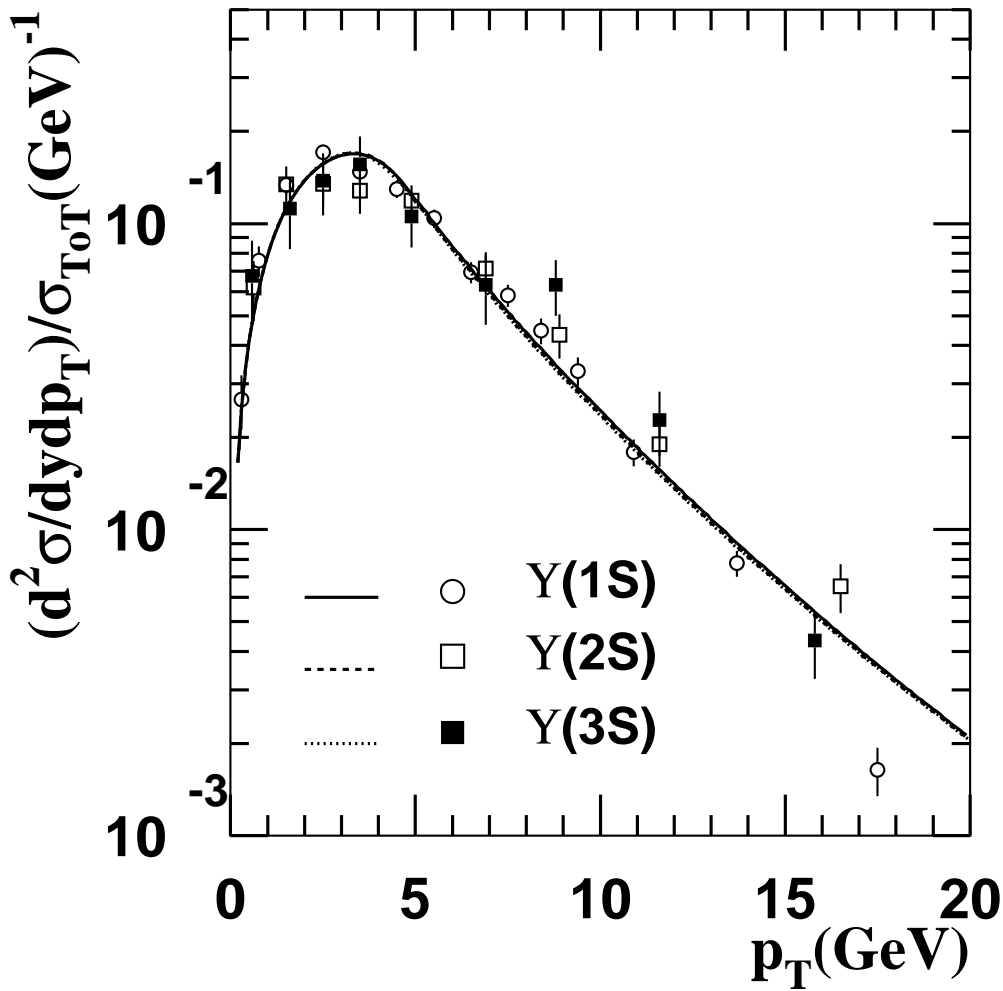}}
\caption{Normalized transverse momentum distributions for $\Upsilon$
  production: $\Upsilon(1S)$ (solid), $\Upsilon(2S)$ (dashed), 
  and $\Upsilon(3S)$ (dotted), along with the 2002  
  CDF data~\cite{Acosta:2001gv}.}  
\label{fig7}
\end{figure}

The essential similarity of the production differential cross sections
for the three $\Upsilon(nS)$ states is illustrated in Fig.~\ref{fig7}.  
Shown are the differential cross sections divided by their respective 
integrals over the range $0 < p_T < 20$~GeV.   
The integrated values are computed from the theoretical cross sections
and used to scale the experimental as well as the theoretical results.  
The three theory curves are practically indistinguishable.  The 
transverse momentum distribution is described well over the full  
range of $p_T$.  Since the curves in Fig.~\ref{fig7} are normalized by 
the integrated cross sections, dependence on the normalization parameters 
$C_{\Upsilon}$ cancels in the ratio.  The shape for $p_T < M_{\Upsilon}/2$ 
is predicted quantitatively.  It reflects the resummation of the 
gluon shower and is independent of parameter choices. The good agreement
with data over the full range in $p_T$ is 
based on the choice of only two adjustable constants, the resummation 
enhancement factor $K_r$ and the matching point $p_{T_M}$. 

\section{Summary and Conclusions}
\label{sec:conclusions}

In this paper we calculate the transverse momentum $p_T$ distribution 
for production of the $\Upsilon$ states in hadronic reactions, applicable 
over the full range of values of $p_T$.  Our starting assumption is that 
the $p_T$ distribution of $\Upsilon$ production may be derived from the 
$p_T$ distribution for the production of a pair $b \bar{b}$ of bottom 
quarks.  We express the differential cross section in terms of a two-step 
factorization procedure.  We justify the validity of an all-orders 
soft-gluon resummation approach to compute the $p_T$ distribution in 
the region of small and intermediate $p_T$ where $p_T < M_{\Upsilon}$.  
Resummation is necessary to deal with the perturbative $1/p_T^2$
singularity and the large logarithmic enhancements
that arise from initial-state gluon showers.  We demonstrate that the  
$p_T$ distribution at low $p_T$ is dominated by the region of small 
impact parameter $b$ and that it may be computed reliably in perturbation 
theory.  We express the cross section at large $p_T$ by the 
${\cal O} (\alpha_s^3)$ lowest-order non-vanishing perturbative 
contribution. Our results are in good agreement with data from $p \bar{p}$ 
interactions at the Fermilab Tevatron collider at center-of-mass 
energy $\sqrt S = 1.8$~TeV, and they confirm that 
the resummable part of the initial-state gluon showers provides the
correct shape of the $p_T$ distribution in the region of small $p_T$.  

An improvement of our calculation in the region $p_T < M_{\Upsilon}$
would require inclusion of the order $\alpha_s$ process-dependent corrections 
associated with the coefficient functions ${\cal C}^{(1)}$ in the CSS
formalism (or equivalently, 
${\cal C}^{(1)}$ and $H_{ij}^{(1)}$ in Eq.~(\ref{cfg-W-H})).  
Based on prior experience~\cite{Berger:2002ut}, we expect that these effects 
will increase the predicted 
normalization of $d\sigma^{\rm resum}/dp_T^2 dy$, and change the shape of 
the $p_T$ distribution somewhat, increasing (decreasing) the spectrum at 
small (large) $p_T$.  A complete calculation of the order $\alpha_s$ 
corrections ${\cal C}^{(1)}_{a\rightarrow i}$ and $H^{(1)}_{ij}$ in
Eq.~(\ref{cfg-W-H}) would provide a better test of QCD predictions.
In the region of large $p_T$, an improved prediction of the normalization 
and shape of the differential cross section would require a formidable 
${\cal O} (\alpha_s^4)$ calculation of $b \bar{b}$ production. 

\begin{figure}[ht]
\centerline{\includegraphics[width=10cm]{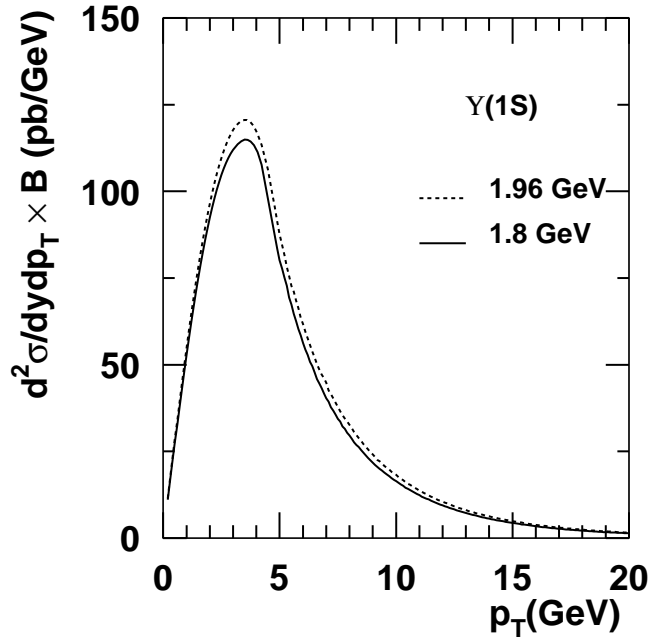}}
\caption{Differential cross sections times leptonic branching 
fractions $B$, evaluated at $y = 0$, as functions of transverse momentum 
for hadronic production of the $\Upsilon(1S)$ at $\sqrt S = 1.8$~TeV (solid 
line) and $1.96$~TeV (dashed line).}
\label{fig8}
\end{figure}

Inclusive production of the $\Upsilon$ states in the central region of 
rapidity at Tevatron energies and above is controlled by partonic subprocesses 
initiated by gluons.  The typical value of the fractional momentum $x$ 
carried by the gluons is determined by the ratio $M_{\Upsilon}/\sqrt S$.  
The growth of the gluon density as $x$ decreases leads to two expected 
changes in our predictions for larger $\sqrt S$.  First, and perhaps 
obvious, the magnitude of the cross section near the peak in, {\em e.g.}, 
Fig.~\ref{fig6} will increase. 
Second, and more subtle is the prediction that the peak location
should shift to a greater value of $p_T$ as $\sqrt S$ grows.  
We use the same parameters as those at $\sqrt{S}=1.8$~TeV, Fig.~\ref{fig6}, 
to compute $\Upsilon(1S)$ production at the Tevatron in run-II at 
$\sqrt{S}=1.96$~TeV; our results are shown in Fig.~\ref{fig8}.  The change 
of $\sqrt S$ from 1.8~TeV to $1.96$~TeV does not produce a marked difference 
in the spectrum, but we expect the shift of the peak in $p_T$ to be about 
$1$~GeV at the LHC energy of $\sqrt S = 14$~TeV.    

Our focus on $\Upsilon$ production may motivate questions about the analogous 
production of the $J/\psi$ states.   The mass of the $J/\psi$ is relatively 
small, meaning that inverse power contributions of the form $1/Q$ are 
potentially as significant as the logarithmic terms $\log(Q)$ that we 
resum.  In addition, the saddle point in the $b$-space distribution moves 
into, or close to the region in which perturbation theory can no longer be 
claimed to dominate the $p_T$ distribution.  

\section*{ACKNOWLEDGMENTS} 
  
Research in the High Energy Physics Division at Argonne is supported 
by the United States Department of Energy, Division of High Energy 
Physics, under Contract W-31-109-ENG-38.  J.~W.~Qiu is supported in part by 
the United States Department of Energy under Grant No. DE-FG02-87ER40371.  
Y.~Wang is supported in part by the United States Department of Energy
under Grant No. DE-FG02-01ER41155.  ELB is grateful to the Kavli Institute 
for Theoretical Physics, Santa Barbara for hospitality during the completion 
of this research. This work was supported in part by the National Science 
Foundation under Grant No. PHY99-07949 and Grant No. PHY00-71027.


\end{document}